\documentclass[preprintnumbers,secnumarabic,amsmath,amssymb,nofootinbib,floatfix]{revtex4}
\usepackage{varioref,exscale,latexsym,amsmath,amssymb}
\usepackage{graphicx}

\usepackage{graphicx}
\usepackage{slashed}
\usepackage{dcolumn}
\usepackage{bm}

\def\beq{\begin{equation}}

\def\eeq{\end{equation}}

\def\beqa{\begin{eqnarray}}

\def\eeqa{\end{eqnarray}}

\begin{document}
\preprint{ACFI-T16-25}

\title{{\bf A conformal model of gravitons }}

\medskip\

\author{ John F. Donoghue${}$}
\email[Email: ]{donoghue@physics.umass.edu}
\affiliation{~\\
Department of Physics,
University of Massachusetts\\
Amherst, MA  01003, USA\\
 }

\begin{abstract}
In the description of general covariance, the vierbein and the Lorentz connection can be treated as independent fundamental fields. With the usual gauge Lagrangian, the Lorentz connection is characterized by an asymptotically free running coupling. When running from high energy, the coupling gets large at a scale which can be called the Planck mass. If the Lorentz connection is confined at that scale, the low energy theory can have the Einstein Lagrangian induced at low energy through dimensional transmutation. However, in general there will be new divergences in such a theory and the Lagrangian basis should be expanded. I construct a conformally invariant model with a larger basis size which potentially may have the same property.
\end{abstract}
\maketitle


\section{Introduction}

We are used to thinking of the Planck scale ($M_p =1.2 \times 10^{19}$~GeV) as a barrier to our physical theories. Indeed, because gravitational quantum fluctuations are of order $E^2/M_p^2$, where $E$ is an energy, they become of order unity at this scale and we appear to lose even the usual concept of spacetime. But
perhaps a simple theory could exist beyond $M_P$, which would be approachable only after passing through a temporarily strong coupled energy region. This paper works towards the construction of a conformally invariant model of gravitation which may be asymptotically free beyond the Planck scale.

An analogy may help describe the physical picture that is being proposed. QCD with massless fermions is classically scale invariant\footnote{When coupled to gravity, it is conformally invariant in the sense described below.}. However, after quantum corrections, its description depends on the energy scale under consideration. With two fermions, it is weakly coupled in {\em both} the ultraviolet and in the infrared, although not in between. The UV description is the well-known story of asymptotic freedom. As we come down from weak coupling at large energies, we enter a strongly coupled regime from $2~{\rm GeV} \to 0.5~{\rm GeV}$ where we cannot calculate perturbatively. However, below
this energy range the theory again becomes weakly coupled with pions as the physical degrees of freedom rather than the original quarks and gluons. At low energies
the interactions are described by the Lagrangian
\begin{equation}\label{chiral}
{\cal L} = \frac{F^2}{4} {\rm Tr} (\partial_\mu U\partial^\mu U^\dagger )     ~~~~~~~~~~~~{\rm with } ~~~~~U= exp\left[\frac{ {i {\mathbf \tau}\cdot {\mathbf \pi}}}{F}\right]
\end{equation}
where $\tau^i$ are the three $SU(2)$ Pauli matrices, $\pi^i$ are the three pions and $F=93$~MeV is the pion decay constant. Here the weak coupling comes from the derivative interaction, such that all amplitudes are of the order of the (Energy)$^2$, and hence become very small at low energies. The pions are Goldstone bosons associated with the chiral symmetry of massless QCD, and both their existence and the structure of the Lagrangian follow from the original symmetry.

Note that there is no trace remaining of the original scale invariance.  The low energy Lagrangian involves a dimensionful parameter $F$, which arises from the scale of QCD - a phenomena commonly called ``dimensional transmutation''. Although the Lagrangian is scale invariant, the running coupling constant involves a scale at which QCD becomes strong. The low energy action involves parameters proportional to this scale. The nonlinear Lagrangian of Eq. \ref{chiral}, technically non-renormalizeable, can be used in
a full effective field theory treatment referred to as chiral perturbation theory to calculate a full quantum theory description of weakly coupled pions.

If we think of a hypothetical civilization living at very low energy, they could have uncovered the chiral Lagrangian from its propagating massless pions and long range forces. With enough precision, they could uncover higher order terms in the chiral Lagrangian. To this civilization, the QCD scale of 1 GeV would have seen as a barrier to their theory because the effective field theory falls apart there. Yet a well-behaved weakly interaction quantum field theory does exist beyond that scale. Perhaps this is similar to our view of the effective field theory of gravity.

The goal of this paper is to discuss whether a similar picture could be developed for gravity with an asymptotically free theory at high energy and a weakly coupled effective field theory at low energy \cite{Donoghue:1994dn}. This will be done in the setting where one treats the vierbein (or tetrad) and the Lorentz connection (or spin connection) as independent fields \cite{confined}\footnote{A different treatment of these as independent fields can be found in \cite{Percacci}}. A simple scale invariant model for which asymptotic freedom is known will be seen to have the ability to generate
the Einstein action via dimensional transmutation. However it is also seen that there should be a more general Lagrangian.
This leads us to propose a more complicated action, but one which is constrained by a stronger symmetry - that of local conformal invariance. This action has several parameters and we discuss the possible outcomes of such a theory at low energy.

\section{Preliminary notation}

This section introduces the players in the construction described in this paper. The physics content starts in the following section.

We will use the vierbein and the Lorentz connection as the fundamental fields. The vierbein is defined via
\begin{equation}
g_{\mu\nu}(x) = \eta_{ab}~ e^a_\mu(x)~ e^b_\nu(x)
\end{equation}
where $\eta_{ab}$ is the flat Minkowski metric. One also defines the inverse metric $g^{\mu\nu}$ and inverse vierbein $e_a^\mu$ with $e_a^\mu e^a_\nu(x)= \delta^\mu_\nu $ and $e_a^\mu e^b_\mu(x)= \delta^b_a $. Latin indices are raised and lowered with $\eta^{ab},~~\eta_{ab}$ and Greek ones with $g^{\mu\nu}(x),~~g_{\mu\nu}(x)$.

In addition to the general coordinate invariance, under which the vierbein transforms as
\begin{equation}
e'^a_\mu = \frac{\partial x^\nu}{\partial x'^\mu}e^a_\nu
\end{equation}
there is extra {\em local} Lorentz symmetry
\begin{equation}
e'^a(x) = \Lambda^a_{~c}(x)~e^c(x)  ~~~~{\rm with } ~~~ \eta_{ab}~\Lambda^a_{~c}(x)~\Lambda^b_{~d}(x)  = \eta_{cd}
\end{equation}

When coupled to fermions, the construction of Utiyama and Kibble \cite{Utiyama, Kibble} also includes the Lorentz connection. The gamma matrices are connected to
derivatives via the vierbein
\begin{equation}
  {\cal L} = \bar{\psi} [i \gamma^a e_a^\mu(x)  \partial_\mu +.....]\psi  \ \ .
\end{equation}
In addition, the fermions transform under the local Lorentz symmetry
\begin{equation}\label{localLor}
\psi \to \psi'(x') = S(x) \psi (x)
\end{equation}
where in matrix notation
\begin{equation}
S(x) =  \exp \left( \frac{-i}{2} J_{ab}\alpha^{ab}(x)\right)
\end{equation}
where $\alpha^{ab}(x) $ is the parameter associated with the local Lorentz transformation $\Lambda$ and
\begin{equation}\label{angs}
  J_{ab} = \frac{\sigma_{ab}}{2}~~~~~~~~~~~{\rm with}~~~  \sigma_{ab} =\frac{i}{2}[\gamma_a, \gamma_b]  \ \ .
\end{equation}
In order that this symmetry be local, we introduce the Lorentz connection as a gauge field $\omega^{ab}_\mu$ and covariant derivative $D_\mu$ with
\begin{equation}
 {\cal L} = \bar{\psi} [i \gamma^a e_a^\mu(x) D_\mu]\psi
\end{equation}
with\footnote{Here I absorb the coupling constant for the Lorentz connection into the field, and it gets reintroduced through the coefficients of the gauge term in the action.}
\begin{equation}
 D_\mu = \partial_\mu -i \frac{J_{ab}}{2}\omega^{ab}_\mu  \equiv \partial_\mu -i  \boldsymbol{\omega}_\mu
\end{equation}
Under the local Lorentz transformation of Eq. \ref{localLor}, the fields transform as
\begin{eqnarray}
  \boldsymbol{\omega}_\mu ' &=& S \boldsymbol{\omega}_\mu S^{-1} -{i}(\partial_\mu S)S^{-1}  \nonumber \\
  e_a^{\mu '} &=& \Lambda_a^{~b} (x) e_b^\mu ~~~~~ ~{\rm with}   ~~~~S^{-1}(x)\gamma^a S(x)\Lambda_a^{~b}(x)= \gamma^b
\end{eqnarray}
This combined with the general coordinate transformation
\begin{eqnarray}
  \mathbf{\omega'}_\mu  &=& \frac{\partial x^\nu}{\partial x'^\mu}\mathbf{\omega}_\nu   \nonumber \\
  e'^\mu_a &=& \frac{\partial x'^\mu}{\partial x'^\nu}e_a^\nu
\end{eqnarray}
define the symmetries of the theory.

The metricity condition for the vierbein is
\begin{equation}
 \nabla_\mu e^a_\nu = 0 = \partial_\mu e^a_\nu + \omega^a_{~b\mu} e^b_\nu - \Gamma_{\mu\nu}^{~~\lambda} e^a_\lambda \ \ .
\end{equation}
Here $\Gamma_{\mu\nu}^\lambda $ is the usual connection defined from the metric
\begin{equation}
\Gamma_{\mu\nu}^{~~\lambda}  = \frac12 g^{\lambda\sigma}\left[\partial_\mu g_{\sigma\nu}+ \partial_\nu g_{\mu\sigma}- \partial_\sigma g_{\mu\nu}\right]
\end{equation}
If metricity is imposed, the Lorentz connection can be eliminated as an independent field. However, we do not impose metricity in this paper, and keep both fields as independent.

\subsection{Varieties of derivatives}

It is often useful to define different notations for various combinations of derivatives and connections.
First let us define the simple partial derivatives:
\begin{equation}
\partial_\mu \equiv \frac{\partial}{\partial x^\mu}  ~~~~~~~~~~\partial_a \equiv e^\mu_a \partial_\mu
\end{equation}
Next, it is often useful to define the derivative which includes only the spin connection:
\begin{equation}
d_\mu \equiv \partial_\mu -i \frac{1}{2} J_{ab} \omega^{ab}_\mu ~~~~~~~~~~d_a \equiv e_a^\mu d_\mu
\end{equation}
This has various forms depending on the object that is being acted on. For a scalar
\begin{equation}
d_\mu \phi = \partial_\mu \phi
\end{equation}
while for a spinor
\begin{equation}
d_\mu \psi = \left[ \partial_\mu -i \frac{1}{2} J_{ab} \omega^{ab}_\mu \right]\psi  ~~~~~~~~{\rm with} ~~~~J_{ab} =\frac12 \sigma_{ab}
\end{equation}
and for a Lorentz vector
\begin{equation}
d_\mu A^a = \partial_\mu A^a + \omega^a_{\mu~b} A^b
\end{equation}
We also define the fully covariant derivative, which involves both $\omega_\mu^{ab}$ and $\Gamma_{\mu\nu}^{~~\lambda}$ in the usual ways. In particular the metricity condition displays this covariant derivative
\begin{equation}
\nabla_\mu e^a_\nu = \partial_\mu e^a_\nu + \omega_{\mu~~b}^a e^b_\nu - \Gamma_{\mu\nu}^{~~\lambda} e^a_\lambda
\end{equation}

\subsection{Field strength tensors}

The definition of various field strength tensors follows from commutators of the covariant derivatives.
The field strength tensor for the spin connection is related to the curvature
\begin{equation}
\left[d_\mu, d_\nu \right] = -i\frac{1}{2}J_{ab} ~R^{ab}_{\mu\nu}
\end{equation}
which yields
\begin{equation}\label{fieldstrength}
R^{ab}_{\mu\nu} = \partial_\mu \omega_\nu^{ab} - \partial_\nu \omega_\mu^{ab} + \left(\omega_{\mu~c}^a~\omega_\nu^{cb} -\omega_{\nu~c}^a~\omega_\mu^{cb}\right)
\end{equation}

There is also a field strength tensor for the verbein, which vanishes if metricity is assumed
\begin{eqnarray}
E^{a}_{\mu\nu} &=& \nabla_\mu e^a_\nu - \nabla_\nu e^a_\mu = d_\mu e^a_\nu - d_\nu e^a_\mu \ \ \nonumber \\
&=& \partial_\mu e^a_\nu +  \omega_{\mu~~b}^a e^b_\nu - \partial_\nu e^a_\mu -  \omega_{\nu~~b}^a e^b_\mu  \ \ .
\end{eqnarray}
This tensor is a torsion tensor, although various authors ascribe various meanings to this term.
There is also the dual of this tensor, defined via
\begin{equation}
\tilde{E}^a_{\alpha\beta} =\frac12 \epsilon_{bcde} e^{b\mu}e^{c\nu}e^d_\alpha e^e_\beta E^a_{\mu\nu}
\end{equation}
We can define the structure constants for the translation group
\begin{equation}\label{Fabc}
  \left[d_a, d_b \right]\psi =\left[F_{ab}^{~~c}~d_c -i \frac{1}{2}e_a^\mu e^\nu_b J_{cd} ~R^{cd}_{\mu\nu}\right] \psi
\end{equation}
which yields
\begin{eqnarray}
F_{ab}^{~~c} &=&  \left(d_a e_b^\mu -d_b e_a^\mu\right) e^c_\mu  \ \  \nonumber \\
&=& e_a^\lambda \left(\partial_\lambda e_b^\mu +\omega_{\lambda b}^{~~d} e_d^\mu\right) e^c_\mu- e_b^\lambda \left(\partial_\lambda e_a^\mu +\omega_{\lambda a}^{~~d} e_d^\mu\right) e^c_\mu
\end{eqnarray}

\section{A scale invariant variation}

Let me first present what might be considered to be the simplest model exhibiting the desired phenomena. It by itself could be considered as a UV completion of General Relativity. It allows a presentation of the basic idea in terms of calculations that many readers will already be familiar with. However, as described below, the presence of fermions is not included and that will lead us to consider a more complete version exhibiting conformal invariance, which will be presented in the following sections.

\subsection{The Lorentz connection and dimensional transmutation}

If we want to treat the spin connection as a gauge field, the simplest action is the usual gauge action
\begin{equation}\label{gauge}
{\cal L} = - \frac{1}{4g^2}  R^{ab}_{\mu\nu} R_{ab}^{\mu\nu}  = - \frac{1}{4g^2} g^{\mu\alpha} g^{\nu\beta} R^{ab}_{\mu\nu} R_{ab\alpha\beta}\ \ .
\end{equation}
Here $g$ is a coupling constant for the Lorentz connection (not to be confused with the determinant of the metric). This Lagrangian has the full general coordinate invariance and local Lorentz invariance. The Lagrangian itself is scale invariant.  As shown in Ref. \cite{confined}, this action also defines an asymptotically free theory. The $\beta$-function is negative and the coupling strength grows as one runs towards the infrared. The energy scale where the perturbative coupling gets large defines what may be called the Planck scale. At this energy scale the dynamics is non-perturbative.  It was suggested in \cite{confined}, that this implies that the spin connection is gapped by being either confined or condensed such that it does not propagate at low energy. I will adopt this hypothesis in what follows. The only essential point is that below this Planck scale, the only active degree of freedom is the vierbein or equivalently the metric.

What would the low energy theory look like in this theory? Dimensional transmutation tells us that it need not be scale invariant. However, it still is invariant under general coordinate invariance and local Lorentz invariance. And it must be described by the metric degree of freedom only, as the spin connection is by assumption not present. This implies that should be described by a general expansion of the action in terms of derivatives of the metric. That is, it must have the form of a general coordinate invariant effective field theory of the metric and the associated curvatures. This would start out as
\begin{equation}
S = \int d^4x \sqrt{-g} \left[-\Lambda - \frac{2}{\kappa^2}R(g) +....\right]
\end{equation}
Here $R(g)$ is the usual scalar curvature constructed from the metric, which is the active low energy field, not the scalar curvature $R(\omega)$ constructed by contracting the indices of the field strength tensor of Eq. \ref{fieldstrength}.
This result is the Einstein-Hilbert action, even if the original theory was scale invariant.

In order to see how this might arise, consider the following calculation. If we consider the one-loop effective action using the Lagrangian of Eq. \ref{gauge}, we can evaluate the result using the heat kernel expansion, where in general for some differential operator ${\cal D}$
\begin{equation}
\det {\cal D} = \exp \left[\int d^4 x {\rm Tr }<x|\ln {\cal D}|x>\right]= -\exp \left[\int d^4 x \int_0^\infty \frac{d\tau}{\tau}{\rm Tr }<x|e^{-\tau {\cal D}}|x>\right]
\end{equation}
where I have dropped an inessential constant in the last step. The heat kernel is defined as
\begin{equation}
H(x,\tau)=  <x|e^{-\tau {\cal D}}|x>
\end{equation}
and has the general expansion in the proper time
\begin{equation}
H(x,\tau) = \frac{i}{(4\pi)^{d/2}}\frac{e^{-m^2\tau}}{\tau^{d/2}}\left[a_0(x) + a_1(x) \tau + a_2(x) \tau^2\right]
\end{equation}
The heat kernel coefficients have been previously calculated for a gauge field loop in the presence of gravity \cite{Seeley}. Inserting factors relevant for the six Lorentz connection fields, this amounts to
\begin{eqnarray}
{\rm Tr}~ a_0 = 24  \ \ \nonumber \\
{\rm Tr}~ a_1(x)  = -2 R(g)
\end{eqnarray}
where $R$ is the usual scalar curvature defined in terms of the metric. If we evaluate the heat kernel integral with a finite proper time cutoff $\tau_0$ one ends up with the action.
\begin{equation}
\Delta {\cal S} = \int d^4x \frac{1}{16\pi^2} \left[  \frac{12}{\tau_0^2} - \frac{2}{\tau_0} R +...\right]
\end{equation}
From this we can see directly how the scalar curvature can be generated through quantum effects. Of course, this is not a ``real'' calculation for this theory, which becomes strongly coupled at low energy and for which the one-loop approximation is surely inadequate. And a proper time cutoff is not appropriate to regularize a scale invariant theory. However, this does display the nature of the quantum corrections, and these will remain true in the strongly coupled theory. If one regularizes the theory by a procedure such as dimensional regularization, which respects the scale invariance of the original theory, and treats the full strongly coupled theory the coefficient in the action will be replaced by a factor proportional the scale of the theory, the Planck scale, via dimensional transmutation.

Unlike the QCD analogy discussed in the introduction, the metric exists in both the high energy and low energy limits of the theory. In the low energy limit there is no need of a separate emergent field, as the metric is able to realize the symmetry of the theory. However, its interactions become ``dressed'' by the strong interactions which occur in the intermediate region, allowing new interactions consistent with the symmetry of the low energy theory.

As an aside, one can note that a very similar ``not a full calculation'' can be performed for the chiral Lagrangian of QCD. Briefly described, if we start from the QCD Lagrangian (with again two massless quarks for simplicity) there is an $SU(2)_l \times SU(2)_R$ chiral symmetry, $\psi_L \to L\psi_L $ and $\psi_R\to R\psi_R $ with $L,~R$ in $SU(2)_{L,R}$. In this case, we can can factor out the chiral coordinates through the field redefinition
\begin{equation}
\psi_L = \xi \Psi_L,  ~~~~\psi_R = \xi^\dagger \Psi_R ~~~~~~{\rm with}  ~~~~~~ \xi \to L\xi V^\dagger = V \xi R^\dagger
\end{equation}
with $V$ being an $SU(2)$ matrix with a vectorial transformation property. This involves the non-linear chiral construction of Callen, Coleman and Wess and Zumino\cite{ccwz}. The Dirac action then becomes
\begin{equation}
{\cal L}_D = \bar{\Psi} i\slashed{D} \Psi
\end{equation}
with
\begin{eqnarray}
D_\mu &=& \partial_\mu +i V_\mu +i A_\mu\gamma_5 \nonumber \\
V_\mu &=& -\frac{i}{2}\left(\xi^\dagger \partial_\mu \xi + \xi\partial_\mu\xi^\dagger\right) \nonumber \\
A_\mu &=& -\frac{i}{2}\left(\xi^\dagger \partial_\mu \xi - \xi\partial_\mu\xi^\dagger\right)   \ \ .
\end{eqnarray}
Integrating out the field $\Psi$ at one loop, for this operator\footnote{The calculation is described in Appendix B of \cite{Donoghue:1992dd}.} the $a_1(x)$ coefficient is given
by
\begin{equation}
{\rm Tr'} ~a_1(x) = -8{ \rm Tr}~ A_\mu A^\mu = 8{ \rm Tr}~ (\partial_\mu U \partial^\mu U^\dagger)
\end{equation}
with $U=\xi^2$. Here the symbol ${\rm Tr'}$ includes a trace over the Dirac indices, while the ${ \rm Tr}$ on the right hand side is only the trace over the flavor indices. Imposing a proper time cutoff, one obtains for the
one loop action
\begin{equation}\label{chiral2}
\Delta {\cal L} = \frac{8}{\tau_0}{\rm Tr} (\partial_\mu U\partial^\mu U^\dagger )
\end{equation}
Again the proper time cutoff is not a real property of scale invariant QCD, and the one loop approximation is not appropriate for a strongly coupled theory. But if we invoke dimensional transmutation to replace the overall coefficient by something proportional to $\Lambda_{QCD}^2$, we identify the general structure of the low energy chiral Lagrangian. In both these cases, the heat kernel is useful in identifying the structure of the induced quantum loop effects.

Returning to gravity, it appears that we have identified a simple, renormalizaeable, asymptotically free gauge theory which yields the Einstein-Hilbert action in the low energy limit. However, the theory is not yet complete. The most telling indication is the $a_2$ coefficient in the heat kernel expansion. If we now include this we find a one loop divergence
\begin{equation}\label{oneloop}
\Delta {\cal L} =\frac{3}{160\pi^2}\frac{2}{d-4} C_{\mu\nu\alpha\beta}C^{\mu\nu\alpha\beta}
\end{equation}
where $C_{\mu\nu\alpha\beta}$ is the Weyl tensor
\begin{eqnarray}
C_{\mu\nu\alpha\beta} &=& {R}_{\mu\nu\alpha\beta} -\frac12 \left( {R}_{\mu\alpha} g_{\nu\beta} - {R}_{\nu\alpha} g_{\mu\beta} - {R}_{\mu\beta} g_{\mu\alpha} + {R}_{\nu\beta} g_{\mu\alpha} \right) \ \ \nonumber \\
&+& \frac{{R(g)}}{6}\left(g_{\mu\alpha} g_{\nu\beta} - e_{\nu\alpha} e_{\mu\beta} \right)
\end{eqnarray}
Here I have switched to dimensional regularization in order to highlight an important feature. This term does represent a true divergence in this scale invariant theory. Dimensional regularization respects the scale invariance, and the result in Eq. \ref{oneloop} is also scale invariant. Of course, in a strongly coupled theories there will be higher order loops which also contribute but the divergence of the form of Eq. \ref{oneloop} is one that is expected on general grounds to be present at each order of perturbation theory.

\subsection{A first conformal model}

In order to renormalize this theory, we then need to include a term involving $C^2$ in the action. For this section then, our basic theory consists of the action
\begin{equation}\label{scaleinvariant}
S_{s.i.} = \int d^4x\sqrt{-g} \left[ - \frac{1}{4g^2}  R^{ab}_{\mu\nu} R_{ab}^{\mu\nu}  - \frac{1}{2\xi^2} C_{\mu\nu\alpha\beta}C^{\mu\nu\alpha\beta} \right]
\end{equation}
The label ${s.i.}$ stands for scale invariant.  Both $g$ and $\xi$ are dimensionless coupling constants, and the action $S_{s.i.}$ contains no dimensionful parameters. In fact, this action has conformal invariance.
This will be discussed in much more detail in the following section, but for the purposes of the present section it implies the invariance under \begin{equation}\label{confg0}
  g_{\mu\nu} \to \Omega^2(x)  g_{\mu\nu}    ~~~~~~~~~ g^{\mu\nu} \to \Omega^{-2}(x)  g^{\mu\nu}
\end{equation}
with $\Omega(x)$ being an arbitrary space-time dependent scaling factor. As is well known and as will be discussed further below, the Weyl term in the action
is invariant under this transformation. Moreover, if we treat a gauge field as invariant under the conformal transformation
\begin{equation}\label{nochange}
\omega_\mu^{ab} \to \omega_\mu^{ab}
\end{equation}
then the gauge term in the action is also invariant. I do want to emphasize here that this simple transformation property of the Lorentz connection will be modified in the presence of fermions. This will be discussed in the following section. However the combined conformal invariance of Eq. \ref{confg0} and Eq. \ref{nochange} is important for the analysis of the action of Eq. \ref{scaleinvariant}. The combined action then is conformally invariant. These are the only two structures that involve only the Lorentz connection and the metric that have this invariance.

\subsection{Power counting}

In the usual effective field theory of general relativity, there is a power counting theorem describing the effects of graviton loops. This is related to Weinberg's power counting theorem in chiral theories  \cite{Weinberg:1978kz}. Specifically, starting from the Einstein-Hilbert action $\sim R$, one loop effects enter at order $R^2$, two loops modify physics at order $R^3$, etc. This can be seen most easily by counting powers of $\kappa^2\sim 1/M_P^2$. With this dimensionful expansion parameter, and massless particles in loops, the powers of $1/M_P^2$ must be compensated for by derivatives (or equivalently of curvatures) in the numerator. This is the hallmark of a so-called non-renormalizeable effective field theory. The loop expansion involves an ever increasing basis. All the divergences are local and can be absorbed into the remormalization of local terms in the most general Lagrangian. High powers of the curvature are unimportant at low energy, although still one needs ever increasing powers of the curvature to deal with higher number of loops.

Matter fields start out similarly. One loop effects of massless matter particles coupled to gravity induces effects at order $R^2$. However, at this stage the power counting differs. If we calculate two loop diagrams of massless matter fields, where all the particles involved in the loops are the matter fields coming from renormalizeable field theories, the results are still at order $R^2$. We do not get an ever increasing set of powers with extra matter fields. This result can also be seen from simple dimensional counting arguments. Because the matter coupling constants are dimensionless, extra loops cannot involve extra factors of curvatures or derivatives. In particular, an important consequence of this result is that all the divergences which follow from only the matter fields of a renormalizeable QFT, treated to all orders, can be absorbed into the counterterms at order $R^2$. So the power counting of pure matter loops is different from graviton loops, in that it stops at order $R^2$.

Now consider a different gravitational theory in which the Einstein-Hilbert action is absent at the fundamental level, and the gravitational action is of order $R^2$, in particular the $C^2$ action found above. This action starts out involving four derivatives. Constructed from such an action, the propagators behave as $1/q^4$ rather than $1/q^2$. This leads to different power counting rules for the gravitons. In particular, all the divergences from graviton loops  also stop at order $R^2$ independent of the number of gravitational loops. Theories where the kinetic terms are of order four derivatives are potentially dangerous, and there is a large literature about the dangers and their possible solutions \cite{Smilga:2005gb, Sezgin:1979zf, Antoniadis:1986tu, Salvio:2015gsi, Bender:2007wu}. I will not adjudicate these here, but see the discussion of Ref. \cite{Donoghue:2017fvm}. However, from the view of pure perturbation theory and power counting, we can describe how such theories behave. Because the coefficient of $C^2$ in the action is dimensionless, the usual expansion of the fields with their conventional normalization $g_{\mu\nu}= \bar{g}_{\mu\nu} + \kappa h_{\mu\nu}$, where $\bar{g}$ is a background metric and $h$ is the quantum fluctuation, starts off at order
\begin{equation}
C^2 \sim  \frac{1}{M_P^2} h \partial^4 h
\end{equation}
where I have omitted all of the tensor indices. The propagator for the gravitational field then behaves as $M_P^2/q^4$ in place of the usual $1/q^2$. The extra power of $M^2_P$ in the numerator exactly compensates for the $\kappa^2\sim 1/M_P^2$ from the couplings of this graviton. Higher loops therefore do not generate higher powers of $\kappa^2\sim 1/M_P^2$ and the results stay at the order of the curvature squared. This result is perhaps more obvious with the use of a non-conventionally normalized quantum field, with $\kappa h_{\mu\nu} \to h_{\mu\nu}$. In this case, $\kappa$ and $M_P$ never appear in the Feynman rules and there are only dimensionless constants. With no extra dimensional factors, the divergences must stay scale invariant, i.e. of order $R^2$. This has been verified in direct calculations at one loop, and will persist at higher loops.

Power counting then tells us that the theory defined by Eq. \ref{scaleinvariant} will only generate divergences in terms which are also scale invariant.

\subsection{Coupled evolution}

Moreover, there is a yet stronger result. The divergence structure of the action of Eq. \ref{scaleinvariant} will be closed if the theory is properly gauge fixed and regularized. That is, the only divergences will go into the renormalization of $g$ and $\xi$. In this section I describe the renormalization group behavior of this coupled theory.

When treated with a regulator that respects the conformal symmetry, and a gauge fixing procedure that does not break the symmetry explicitly, the divergence structure will also respect the symmetry and hence will renormalize only the couplings that occur in the action. Moreover, the Weyl term only involves the metric, and hence can only renormalize itself, and not the action for the Lorentz connection. On the other hand, the Lorentz connection action involves both the connection and the metric and hence renormalizes both terms in the action. This is born out in direct calculations.

The coupling constants of this theory run. In general, the running will be coupled, $\beta(g)= f_g(g,\xi)$ $\beta(\xi)= f_\xi(g,\xi)$. However at one loop order the running decouples and each runs separately. Both are asymptotically free at one-loop order. In the case of the Weyl term, it forms a perturbatively renormalizeable theory and the beta function is well-known \cite{Stelle:1976gc, Barth:1983hb, Fradkin:1981iu, bos, Antoniadis:1992xu, deBerredoPeixoto:2003pj, Avramidi:1985ki, Avramidi:1986mj}. Treating the Lorentz connection as an independent field, the running of the connection coupling was calculated in Ref. \cite{confined}. The effects of matter fields on the running of on the Weyl coupling can be found from the above divergence. Overall we have
\begin{eqnarray}\label{beta}
\beta(g) &=& -\frac{22}{3\pi^2} g^3 \nonumber \\
\beta(\xi)&=& - \frac{199}{480\pi^2} \xi^3 - \frac{3}{80\pi^2} \xi^3
 \ \ .
\end{eqnarray}
where the first term in $\beta(\xi)$ comes from graviton loops and the second is from the Lorentz connection, and is dominated by the graviton
contribution\cite{bos}. Note also that for a given coupling, the running of $g$ is faster than that of $\xi$.

The running couplings define scales through dimensional transmutation. The assumption of the present work is that the running of the Lorentz connection is most important and will define the Planck scale, where its interactions become strong, and below which energy the Lorentz connection does not propagate. The running of the Weyl term will be modified below the Planck scale, and most importantly becomes sub-dominant to the induced Einstein-Hilbert action. Other authors who have studied the running of the Weyl term in the case where metricity is imposed include
Smilga \cite{Smilga:1982se}, Holdom and Ren \cite{Holdom:2015kbf}, Salvio and Strumia \cite{Salvio:2014soa} and Einhorn and Jones \cite{Einhorn:2014gfa}, and dimensional transmutation is also relevant for this case.

\subsection{Other gauge theories}

In fact, using the Lorentz connection as the gauge field is not required for this construction. Any gauge theory will also induce a similar running in the Weyl coupling. If we simply substitute $R^{ab}_{\mu\nu}$ by any gauge field strength tensor $F^i_{\mu\nu}$, with gauge coupling $g_g$, the one-loop running of the Weyl coupling will be
\begin{equation}
\beta (\xi)= - \frac{1}{480\pi^2} \xi^3 [199 +3D]
\end{equation}
where $D$ is the dimension of the group (i.e. the number of gage fields in the adjoint representation, $D=N^2-1$ for $SU(N)$ or $D= N(N-1)/2$ for $O(N)$).

All gauge fields will then contribute to the running of the Weyl coupling. Under this construction, the Planck scale will be set by gauge theory with the largest intrinsic scale in the running of its gauge coupling. This description of Yang-Mills driven gravity is discussed in more detail in Ref. \cite{Donoghue:2017fvm}.

\subsection{Proceeding further}

The model of this section is closed and self-contained, presenting the coupled evolution of a simple action for the metric and the Lorentz connection. It has been argued that it induces the Einstein-Hilbert action as a finite term if confinement occurs. It is then an asymptotically free model for induced general relativity deserves further study on its own.

However, given our starting point, there is also an unsatisfactory aspect to this model. We motivated treating the Lorentz connection as a gauge field by considering its couplings to fermions, along with that of the vierbein.  If one treats these as independent fields, and computes the effect of a fermion loop, there will be a divergence that is proportional to $C^2$ but also another that is distinct from the field strength term of the Lorentz connection\cite{confined}. With fermions, we need to expand the operator basis.

However, the results that we have seen so far suggest a particular pathway. The Lagrangian with the Weyl tensor squared is conformally invariant. So is the effect of the fermion loop. (Let me defer a discussion of this symmetry to the following section.) This suggests that one might want to keep local conformal invariance as the defining symmetry
\cite{Mannheim:2011ds, Ivanov:1981wn, Wheeler:2013ora, Iorio:1996ad, Englert:1976ep, 'THooft:2015skl, Hooft:2015rdz, Blas:2011ac, Karananas:2016ltn}. The divergence structure would then respect this symmetry, and renormalization would be closed with a general conformally invariant action.

The conformal properties of the Lorentz connection have not been previously elucidated in a framework in which one includes only the vierbein, fermions and Lorentz connection as active fields, and this is the goal of the next section. Note that there are well-known ways to construct conformally invariant actions which introducing involve more fields, with different transformation properties than described below. See for example Refs. \cite{Karananas:2016ltn} and \cite{Freedman:1976xh}. Here we are using only the vierbein and the Lorentz connection.

\section{Conformal invariance and the Lorentz connection}

Local conformal symmetry is a more powerful symmetry than scale invariance. In the gravitational sector, it is defined by
\begin{equation}\label{confg}
  g_{\mu\nu} \to \Omega^2(x)  g_{\mu\nu}    ~~~~~~~~~ g^{\mu\nu} \to \Omega^{-2}(x)  g^{\mu\nu}
\end{equation}
with $\Omega{x}$ being an arbitrary space-time dependent scaling factor.  It is sometimes useful to parameterize this as
\begin{equation}\label{sigma}
\Omega^2(x)  = e^{2\sigma(x)}
\end{equation}
and this allows the notational simplification
\begin{equation}
\Omega^{-1}(x) \partial_\mu \Omega(x) = \partial_\mu \sigma  \ \ .
\end{equation}
Of course the vierbein then transforms as
\begin{equation}\label{confe}
  e_{\mu}^a \to \Omega(x)  e_{\mu}^a    ~~~~~~~~~ e^{\mu}_a \to \Omega^{-1}(x)  e^{\mu}_a
\end{equation}
and we have
\begin{equation}\label{detg}
 \sqrt{-g} \to \Omega^4(x)   \sqrt{-g}  \ \ .
\end{equation}

\subsection{Transformation of the Lorentz connection}

The massless Dirac action can be made conformally invariant by the appropriate transformation of the Lorentz connection. Using
\begin{equation}
S_D = \int d^4x \sqrt{-g} \bar{\psi}\left[ i\gamma^a e^\mu_a (\partial_\mu -  i\frac{J_{ab}}{2}\omega^{ab}_\mu ) \right]\psi
\end{equation}
we can make this invariant under
\begin{equation}
\psi\to \Omega^{-3/2} \psi
\end{equation}
if the Lorentz connection transforms as
\begin{equation}\label{connectiontrans}
\omega^{ab}_\mu \to \omega^{ab}_\mu +(e^a_\mu \partial^b \sigma -e^b_\mu \partial^a \sigma )  \ \ .
\end{equation}
By construction, this is the same condition as if metricity were to be assumed. Note that this is different from the behavior of a usual gauge field, which is normally treated as invariant under a conformal transformation.

The metric connection transforms as
\begin{equation}
\Gamma_{\mu\nu}^{~~\lambda} \to \Gamma_{\mu\nu}^{~~\lambda} +\left(\partial_\mu \sigma \delta^\lambda_\nu +\partial_\nu \sigma \delta^\lambda_\mu -g_{\mu\nu} \partial^\lambda \sigma \right)
\end{equation}

\subsection{A Weyl tensor for the Lorentz connection}

None of the curvatures have a simple conformal transformation, but the Weyl tensor
does transform covariantly
\begin{equation}
C_{\mu\nu\alpha\beta} \to \Omega^2 C_{\mu\nu\alpha\beta}
\end{equation}
leaving the Weyl action
\begin{equation}
S_W = \int d^4x \sqrt{-g} ~-\frac{1}{2\xi} ~C_{\mu\nu\alpha\beta}  C^{\mu\nu\alpha\beta}
\end{equation}
as conformally invariant. This is the unique conformally invariant possibility constructed purely from the metric.

However, the equivalent to the Weyl tensor is not conformally invariant when constructed from Lorentz connection using $R^{ab}_{\mu\nu}$.
I find that under a conformal transformation
\begin{eqnarray}\label{confR}
\delta R^{ab}_{\mu\nu} &=& (d_\mu \partial^b \sigma ) e^a_\nu - (d_\nu \partial^b \sigma ) e^a_\mu \ \ \nonumber \\
&-& (d_\mu \partial^a \sigma ) e^b_\nu +(d_\nu \partial^a \sigma ) e^b_\mu \ \ \nonumber \\
&+& \partial^b\sigma  E^a_{\mu\nu} - \partial^a\sigma  E^b_{\mu\nu}
\end{eqnarray}
Here the first two rows are expected terms and these contributions cancel when forming the equivalent of the Weyl tensor using $R^{ab}_{\mu\nu}$. However the last row does not cancel in this construction. Note that the last row vanishes when metricity is assumed, so the new feature in the conformal tensor is associated with the lack of metricity. Let me refer to the last row as `the extra term'.

In order to construct an invariant object we need to introduce new variables in addition.
We start by noting that the metricity condition transforms covariantly
\begin{equation}\label{conformalmetricity}
\nabla_\mu e^a_{\nu} \to \Omega \nabla_\mu e^a_{\nu}  \ \ .
\end{equation}
This implies that
vierbein tensor also transforms simply
\begin{equation}
E^a_{\mu\nu} \to \Omega E^a_{\mu\nu} ~~,~~~~~~~~~~~~~~ \tilde{E}^a_{\mu\nu} \to \Omega \tilde{E}^a_{\mu\nu}\ \ .
\end{equation}
Among other objects, a useful relation involves
\begin{equation}
F_{ab}^{~~c} \to \Omega^{-1}\left[F_{ab}^{~~c}  +2(\partial_a\sigma \delta^c_b-\partial_b \sigma \delta^c_a) \right]
\end{equation}

These relations allow the construction of a new conformally invariant tensor for the Lorentz connection. If we define a modified curvature
\begin{equation}
\bar{R}^{ab}_{\mu\nu} = R^{ab}_{\mu\nu}  +\frac12 F^{ab}_{~~c} E^c_{\mu\nu}
\end{equation}
then the conformal transformation no longer has the extra term of Eq. \ref{confR}. We can then form the conformally invariant tensor
\begin{eqnarray}
D^{ab}_{\mu\nu} &=& \bar{R}^{ab}_{\mu\nu} -\frac12 \left( \bar{R}^{a}_{\mu} e^b_\nu - \bar{R}^{a}_{\nu} e^b_\mu - \bar{R}^{b}_{\mu} e^a_\mu + \bar{R}^{b}_{\nu} e^a_\mu \right) \ \ \nonumber \\
&+& \frac{\bar{R}}{6}\left(e^a_\mu e^b_\nu - e^a_\nu e^b_\mu \right)
\end{eqnarray}
As stated, this is conformally invariant
\begin{equation}
D^{ab}_{\mu\nu} \to D^{ab}_{\mu\nu}
\end{equation}
which means that the corresponding action
\begin{equation}
S_D = \int d^4x \sqrt{-g}~ D^{ab}_{\mu\nu} D_{ab}^{\mu\nu}
\end{equation}
is also conformally invariant.

\subsection{Conformally invariant spacetime vectors}

When the metricity condition does not vanish there are greater opportunities for conformally invariant vectors. Starting from the conformal transformation of the metricity condition, Eq. \ref{conformalmetricity}, we note that the following vectors are conformally invariant
\begin{equation}
V^{ab}_{1\mu} = e^{a\nu} \nabla_\mu e^b_\nu = V^{[ab]}_{1\mu}
\end{equation}
\begin{equation}
V^{ab}_{2\mu} = e^a_{\mu} \nabla^\nu e_\nu^b
\end{equation}
\begin{equation}
V^{ab}_{3\mu} = e^{a\nu}E^b_{\mu\nu}
\end{equation}
\begin{equation}
V^{ab}_{4\mu} = e^{a\nu}\tilde{E}^b_{\mu\nu}
\end{equation}
along with the contractions
\begin{equation}
V_{1\mu} = 0 = \eta_{ab} V^{ab}_{1\mu}
\end{equation}
\begin{equation}
V_{2\mu} = e_{b\mu} \nabla^\nu e_\nu^b = \eta_{ab} V^{ab}_{2\mu}
\end{equation}
\begin{equation}
V_{3\mu} = e_b^{\nu}E^b_{\mu\nu} = \eta_{ab} V^{ab}_{3\mu}
\end{equation}
\begin{equation}
V_{4\mu} = e_b^{\nu}\tilde{E}^b_{\mu\nu} = \eta_{ab} V^{ab}_{4\mu}
\end{equation}
All of these are conformally invariant and spacetime vectors. The combinations with un-contracted Lorentz indices are Lorentz tensors. The vector $V_4$ has the opposite parity from the other vectors.

For the contracted vectors, with no free Lorentz indices, we can then form conformally invariant field strength tensors
\begin{equation}
V_{i\mu\nu} = \partial_\mu V_{i\nu} - \partial_\nu V_{i\mu}        ~~~~~~~(i~=~2,~3,~4)
\end{equation}

\section{The conformally invariant basis}

We now have the tools to construct a conformally invarinat action from the vierbein and the Lorentz connection in the situation where the Lorentz connection transforms as in Eq. \ref{connectiontrans}. Fortunately or unfortunately, there are many possible conformal invariants.

\subsection{The field strengths}

We start with those terms which yield bilinears in the fields. These are the basic actions which define the kinetic energies and the propagators. The field strength tensors $C_{\mu\nu\alpha\beta}$, $D_{\mu\nu}^{ab}$ and $W_{ab}$ all start of linear in their field variables. Conformally invariant combinations of these are then
\begin{eqnarray}
{\cal L}_2 &=& -\frac{1}{4g_1^2} D_{\mu\nu}^{ab}D^{\mu\nu}_{ab}  - \frac{1}{2\xi^2} C_{\mu\nu\alpha\beta}C^{\mu\nu\alpha\beta}  \ \nonumber  \\
&+&  + \alpha C_{\mu\nu\alpha\beta} e^\alpha_a e^\beta_b D^{ab\mu\nu}  \ \ .
\end{eqnarray}
These have been written under the assumption of parity invariance.

\subsection{A designer propagator}

The field strength tensors of the previous section start off bi-linear in the fields, and can be used to define propagators. However, there can be other conformally invariant terms which also are bi-linear in the fields. These all vanish if metricity is assumed, but remain when treating the fields as independent. Because these terms come with coefficients can be adjusted, one can use these to modify the forms of the propagators if desired.

I have described field strengths composed of conformally invariant vector combination of the tetrad and Lorentz connection. These can be combined to produce conformally invariant actions. Assuming parity conservation, we have the terms
\begin{equation}
S_2=\int d^4x -\frac14\left[\gamma_2 V_{2\mu\nu}V^{\mu\nu}_2 + \gamma_3 V_{3\mu\nu}V^{\mu\nu}_3+ \gamma_{23} V_{2\mu\nu}V^{\mu\nu }_3+ \gamma_4 V_{4\mu\nu}V^{\mu\nu}_4  \right]
\end{equation}
Besides the usual diagonal contributions to propagators, each of these contain off-diagonal mixing between the vierbein and the Lorentz connection.
The propagators of the tetrad and Lorentz connection form a matrix. The diagonal elements for the Lorentz connection are all of order two derivatives. For the tetrad all the diagonal elements are of order four derivatives. The off-diagonal mixing terms involve 3 derivatives.

\subsection{Interaction terms}

There are also invariants that start off at third order in the field variables. The all include only one of the field tensors  $C_{\mu\nu\alpha\beta}$, $D_{\mu\nu}^{ab}$ and $V_{i\mu\nu}$, with the remaining ingredients coming from two powers of the metricity condition confomal vectors. Here we find
\begin{eqnarray}
{\cal L}_3 &=& \sum\limits_{ij}C^{\mu\nu\alpha\beta}\left[a_{ij}e^\mu_ee^\alpha_f\eta_{cd}+ b_{ij}e^\mu_ce^\alpha_f\eta_{de}\right] V^{cd}_{i\nu}V^{ef}_{j\beta}
 \ \nonumber  \\
&+& \sum\limits_{ij}D^{\mu\nu}_{ab}\left[c_{ij}\delta^a_{e}\delta^b_{f}\eta_{cd}+ d_{ij}\delta^a_{c}\delta^b_{f}\eta_{de}\right] V^{cd}_{i\mu}V^{ef}_{j\nu} \ \nonumber\\
&+& \sum\limits_{ij} V^{\mu\nu}_{i}\left[f_{ijk}\eta_{ef}\eta_{cd} + g_{ijk}\eta_{cf}\eta_{de}\right] V_{j\mu}^{cd}V_{k\nu}^{ef}
\end{eqnarray}
Some of the terms in the sum vanish by symmetry considerations, and cross-terms between $V_4$ and $V_{1,2,3}$ are forbidden by parity.

Finally there are interaction terms which start off at fourth order in the fields. These are constructed from variants of the conformal vectors
\begin{equation}
{\cal L}_4 = -\sum\limits_{ijkl}{\rm Perms}_{ab..h}~ \lambda_{ijkl}~ V_{i\mu}^{ab}V_j^{cd\mu} V_{k\nu}^{ef}V_l^{gh\nu}
\end{equation}
where the permutations are taken over the possible contract of the Lorentz indices (with different couplings $\lambda$ for different permutations). All of the terms in this section vanish if metricity is imposed.

\section{Comments on the second conformal model}

We have seen that conformal symmetry has constrained the action. However, there are still multiple parameters. This can be both a difficulty and an opportunity. As a difficulty, it is clear that to fully analyse the general model with require an exhaustive exploration of the parameter space. However, the flexibility of the model may prove beneficial. Admittedly, with a quartic action for the metric and a non-compact gauge group, the model has potential for pathologies. There may be special ranges of the parameters which help solve these problems. For example, the quartic terms in the action help prevent large excursions in the Lorentz connection because they contain positive definite terms in the Hamiltonian Hamiltonian for an appropriate choice of the signs of the coupling constants.

The first requirement for analysing the model is to fine a useful gauge fixing term which does not explicitly violate conformal invariance. Analyses of conformal gravity at one loop have using gauge fixing which explicitly breaks the symmetry leads to divergences which also do not respect the symmetry. This will be addressed in future work.

\subsection{Unimodular gravity, the conformal anomaly and the cosmological constant}

Any metric can be factored into a conformal factor and a unimodular metric
\begin{equation}
g_{\mu\nu}(x) =\Omega^2(x) \hat{g}_{\mu\nu} (x)  ~~~~~{\rm with}  ~~~~~\det (\hat{g}_{\mu\nu}) = -1
\end{equation}
The Lagrangian constructed above is invariant under the conformal transformation, and hence does not depend at all on the conformal factor $\Omega(x)$. The unimodular nature of the action can be seen by direct construction. The simplest case to see this directly is that of the gauge action for the Lorentz connection. Here if one provides a field redefinition $\sqrt[4]{-g}g^{\mu\nu} = \hat{g}^{\mu\nu}$,
\begin{equation}\label{uni}
  \int d^4x~\sqrt{-g} g^{\mu\alpha} g^{\nu\beta} R^{ab}_{\mu\nu}R_{\alpha\beta a b} = \int d^4x~ \hat{g}^{\mu\alpha} \hat{g}^{\nu\beta} R^{ab}_{\mu\nu}R_{\alpha\beta a b}
\end{equation}
and only the unimodular field $\hat{g}^{\mu\nu}$ appears in the action. That this can be done for all the terms in the action follows from the conformal construction. If we choose $\Omega^2 =\sqrt[4]{-g}$, such that $g^{\mu\nu} = \sqrt[4]{-g}\hat{g}^{\mu\nu}$, $\hat{g}^{\mu\nu}$ will be unimodular, and because of the conformal invariance the full action can be written in terms of $\hat{g}^{\mu\nu}$.

The fate of this theory then depends on the path integral measure. If the measure is only invariant under unimodular transformations, then
the low energy theory is that of unimodular gravity, which is classically equivalent to General Relativity. However if the fields in the measure are taken to transform under the full general coordinate transformations, then there can be a conformal anomaly. Even if the Lagrangian is invariant under conformal transformations, the measure is not. This leads extra terms in the effective action.

The conformal anomaly does not follow from any local Lagrangian. Instead it is described by finite but non-local terms in the effective action. There is a large literature on this topic, and there remains a disagreement whether the nonlocal action is described solely by terms which
behave\cite{Deser:1976yx, Erdmenger:1996yc} as $C_{\mu\nu\alpha\beta} \log \Box C^{\mu\nu\alpha\beta}$ or whether the nonlocality goes as $1/\Box^2$ as in the Riegert action\cite{Riegert}. The resolution of this debate is not relevant for the present construction. However, let me comment that with the most standard definiton of $\log \Box$, it appears that both types of terms are needed\cite{Donoghue:2015xla, Barvinsky:1994cg}, although it is possible that one may find an alternate covariant definition of $\log \Box$ that combines both effects. In the enlarged model presented here, it is also possible that some parameters will lead to renomalization group flow to an IR fixed point at which the conformal anomaly vanishes.

Unimodular gravity derived from the scalar curvature has all the same classical predictions as Einstein general relativity. However, the cosmological constant enters the theory in a different way\cite{Unruh:1988in, Ng:1990xz, Weinberg:1988cp}, as an integration constant for a constraint on the equations of motion. The metric no longer couples to the constant term - the vacuum energy - which appears in the action. While there still is a cosmological constant in the equations of motion, it no longer is the measure of the energy of the vacuum, but rather is a feature of the initial conditions of a particular solution. This is important as it allows us to decouple the vacuum energy from the problem of the cosmological constant. At this stage we do not have a theory of cosmology within this conformal model, but some general features could be assumed.  Because $\Lambda$ is set by an initial condition,
that condition could be set during the conformal phase of the universe. At this stage there is no explicit scale in the theory and all
contributions satisfy $T_\mu^\mu=0 $ and the integration constant could vanish simply from the lack of any dimensionful scale at this energy. It is then plausible that the initial condition should set the integration constant to zero.

\subsection{Applying metricity in reverse}

The following is a comment somewhat outside the primary development of this paper. It is prompted by the observation that the Lorentz connection
is a more natural variable for a fundamental theory. This suggests that an alternate possibility is to use the metricity condition to eliminate
the vierbein, writing it in terms of the Lorentz connection. Krasnov \cite{Krasnov:2011pp} has succeeded in doing this for the usual Einstein-Hilbert action, with a rather complicated looking action for the connection, but with significant success in extracting amplitudes \cite{Delfino:2012aj}. For a related attempt involving the Weyl action, see Ref. \cite{Basile:2015jjd}.

In our case, this would initially reduce the complicated action to a single term, which
can be taken to be
\begin{equation}
S_D = \int d^4x \sqrt{-g}~ D^{ab}_{\mu\nu} D_{ab}^{\mu\nu}
\end{equation}
with $D^{ab}_{\mu\nu}$ formed using $R^{ab}_{\mu\nu}$ instead of $\bar{R}^{ab}_{\mu\nu}$. Of course, the vierbein or the metric appear implicitly in this equation in connecting the spacetime indices, and the must be solved for in terms of the Lorentz connection.

Of course, used in this way the metricity condition is a very non-linear constraint. In usual general relativity, the
inverse metric $g^{\mu\nu}$ is also defined by a non-linear constraint from the fundamental field $g_{\mu\nu}$ by requiring it to be the inverse
of the metric. This is a local constraint. The metricity constraint used to write the Lorentz connection in terms of derivatives of the metric is
also a local constraint. In order to eliminate the metric as an independent variable it needs to be written in terms of derivatives of the connection. The schematic procedure is described by Krasnov and collaborators in Ref. \cite{Herfray:2015fpa}.

It would be very messy to write the action explicitly in terms of the Lorentz connection because the
result would be very non-linear. However, this is
only somewhat different than the usual non-linearity of general relativity, where the inverse metric is defined by a constraint from the real metric field, and it is difficult to write the action only in terms of the metric itself. It would be interesting to explicitly attempt such a construction.

\section{Summary}

This paper presents two models for the gravitational interactions, treating the Lorentz connection and the vierbein as separate fields. If the Lorentz connection is confined or otherwise removed from the spectrum, as in the initial scale invariant variation of Sec. 3, then at low energy the symmetry must be carried entirely by the vierbein. It has been argued that, through dimensional transmutation, this will generate the Einstein action from one that was originally scale invariant. The metric can survive to low energy, although as a field it is dressed by the strong interactions with the Lorentz connection. This model deserves more investigation as a model for gravitons, although without fermions.

Perhaps a more satisfying model is obtained by imposing local conformal symmetry including fermions, leading to the transformation of the Lorentz connection given in Eq. \ref{connectiontrans}. Conformally invariant combinations of fields have been constructed and the result is a rich structure. The model has not yet been analysed fully. For some values of the parameters it is hoped that the model can also be asymptotically free and still describe general relativity at low energies. The presence of mixing terms between the Lorentz connection and the vierbein
may prove useful in allowing the vierbein sector to be well defined despite having contributions to the action that involve four derivatives.

The action for this model has many terms, with coefficients which are in principle separate. This is both a difficulty and an advantage. The difficulty is that of fully analysing the system. Having a large parameter space is calculational difficult and will take a sustained effort to explore thoroughly. I will not do that here. However, there are potentially important positive features. There are obvious issues on which the model may fail, and the extended space may circumvent those pitfalls for certain values of the parameters. For example, any given Lagrangian may have an unstable direction due to the feature that the gauge group is non-compact. However, the addition of an extra Lagrangian can be use to stablize the system.

\section*{Acknowledgements} I would like to thank Eugene Golowich, Leandro Bevilaqua, Ted Jacobson, Michael Endres, Renate Loll, Pierre Ramond, Guido Martinelli, Alberto Salvio, Kirill Krasnov and Martin Luscher for useful conversations about this topic. This work has been supported in part by the National Science Foundation under grants NSF PHY15-20292 and NSF PHY12-25915.


\begin{thebibliography}{99}


\bibitem{Donoghue:1994dn}
  J.~F.~Donoghue,
  ``General Relativity As An Effective Field Theory: The Leading Quantum
  Corrections,''
  Phys.\ Rev.\  D {\bf 50}, 3874 (1994)
  [arXiv:gr-qc/9405057].


\bibitem{confined}
  J.~F.~Donoghue,
  ``Is the spin connection confined or condensed?,''
  arXiv:1609.03523 [hep-th].

\bibitem{Percacci}
  R.~Percacci,
  ``Gravity from a Particle Physicists' perspective,''
  PoS ISFTG {\bf }, 011 (2009)
  [arXiv:0910.5167 [hep-th]].

\bibitem{Utiyama}
  R.~Utiyama,
  ``Invariant theoretical interpretation of interaction,''
  Phys.\ Rev.\  {\bf 101}, 1597 (1956).


\bibitem{Kibble}
  T.~W.~B.~Kibble,
  ``Lorentz invariance and the gravitational field,''
  J.\ Math.\ Phys.\  {\bf 2}, 212 (1961).




\bibitem{Seeley} R.T. Seeley, Proc. Symp. Pure Math., Amer.Math.Soc. 10, 288 (1967).\\
B.DeWitt, {\it Dynamical Theory of Groups and Fields}. (Gordon and Breach. New York 1965).

\bibitem{ccwz}
  C.~G.~Callan, Jr., S.~R.~Coleman, J.~Wess and B.~Zumino,
  ``Structure of phenomenological Lagrangians. 2.,''
  Phys.\ Rev.\  {\bf 177}, 2247 (1969).
  doi:10.1103/PhysRev.177.2247


\bibitem{Donoghue:1992dd}
  J.~F.~Donoghue, E.~Golowich and B.~R.~Holstein,
  {\it Dynamics of the standard model},
  Camb.\ Monogr.\ Part.\ Phys.\ Nucl.\ Phys.\ Cosmol.\  {\bf 2}, 1 (1992).




\bibitem{Weinberg:1978kz}
  S.~Weinberg,
  ``Phenomenological Lagrangians,''
  Physica A {\bf 96}, 327 (1979).



\bibitem{Smilga:2005gb}
  A.~V.~Smilga,
  ``Ghost-free higher-derivative theory,''
  Phys.\ Lett.\ B {\bf 632}, 433 (2006)
  doi:10.1016/j.physletb.2005.10.014
  [hep-th/0503213].
\bibitem{Sezgin:1979zf}
  E.~Sezgin and P.~van Nieuwenhuizen,
  ``New Ghost Free Gravity Lagrangians with Propagating Torsion,''
  Phys.\ Rev.\ D {\bf 21}, 3269 (1980).
  doi:10.1103/PhysRevD.21.3269
\bibitem{Antoniadis:1986tu}
  I.~Antoniadis and E.~T.~Tomboulis,
  ``Gauge Invariance and Unitarity in Higher Derivative Quantum Gravity,''
  Phys.\ Rev.\ D {\bf 33}, 2756 (1986).
  doi:10.1103/PhysRevD.33.2756

\bibitem{Salvio:2015gsi}
  A.~Salvio and A.~Strumia,
  ``Quantum mechanics of 4-derivative theories,''
  Eur.\ Phys.\ J.\ C {\bf 76}, no. 4, 227 (2016)
  doi:10.1140/epjc/s10052-016-4079-8
  [arXiv:1512.01237 [hep-th]].
\bibitem{Bender:2007wu}
  C.~M.~Bender and P.~D.~Mannheim,
  ``No-ghost theorem for the fourth-order derivative Pais-Uhlenbeck oscillator model,''
  Phys.\ Rev.\ Lett.\  {\bf 100}, 110402 (2008)
  doi:10.1103/PhysRevLett.100.110402
  [arXiv:0706.0207 [hep-th]].

\bibitem{Donoghue:2017fvm}
  J.~F.~Donoghue,
  ``Quartic propagators, negative norms and the physical spectrum,''
  arXiv:1704.01533 [hep-th].

\bibitem{Stelle:1976gc}
  K.~S.~Stelle,
  ``Renormalization of Higher Derivative Quantum Gravity,''
  Phys.\ Rev.\ D {\bf 16}, 953 (1977).
  doi:10.1103/PhysRevD.16.953

\bibitem{Barth:1983hb}
  N.~H.~Barth and S.~M.~Christensen,
  ``Quantizing Fourth Order Gravity Theories. 1. The Functional Integral,''
  Phys.\ Rev.\ D {\bf 28}, 1876 (1983).
  doi:10.1103/PhysRevD.28.1876

\bibitem{Fradkin:1981iu}
  E.~S.~Fradkin and A.~A.~Tseytlin,
  ``Renormalizable asymptotically free quantum theory of gravity,''
  Nucl.\ Phys.\ B {\bf 201}, 469 (1982).
  doi:10.1016/0550-3213(82)90444-8

\bibitem{bos}
I.~L.~Buchbinder, S.~D.~Odintsov and I.~L.~Shapiro,
  ``Effective action in quantum gravity,''
  Bristol, UK: IOP (1992)

\bibitem{Antoniadis:1992xu}
  I.~Antoniadis, P.~O.~Mazur and E.~Mottola,
  ``Conformal symmetry and central charges in four-dimensions,''
  Nucl.\ Phys.\ B {\bf 388}, 627 (1992)
  doi:10.1016/0550-3213(92)90557-R
  [hep-th/9205015].
\bibitem{deBerredoPeixoto:2003pj}
  G.~de Berredo-Peixoto and I.~L.~Shapiro,
  ``Conformal quantum gravity with the Gauss-Bonnet term,''
  Phys.\ Rev.\ D {\bf 70}, 044024 (2004)
  doi:10.1103/PhysRevD.70.044024
  [hep-th/0307030].


\bibitem{Avramidi:1985ki}
  I.~G.~Avramidi and A.~O.~Barvinsky,
  ``Asymptotic Freedom In Higher Derivative Quantum Gravity,''
  Phys.\ Lett.\ B {\bf 159}, 269 (1985).
  doi:10.1016/0370-2693(85)90248-5
\bibitem{Avramidi:1986mj}
  I.~G.~Avramidi,
  ``Covariant methods for the calculation of the effective action in quantum field theory and investigation of higher derivative quantum gravity,''
  hep-th/9510140.

\bibitem{Smilga:1982se}
  A.~V.~Smilga,
  ``Spontaneous generation of the Newton constant in the renormalizable gravity theory,''
  IN *ZVENIGOROD 1982, PROCEEDINGS, GROUP THEORETICAL METHODS IN PHYSICS, VOL. 2* 73-77.
  [arXiv:1406.5613 [hep-th]].

\bibitem{Holdom:2015kbf}
  B.~Holdom and J.~Ren,
  ``QCD analogy for quantum gravity,''
  Phys.\ Rev.\ D {\bf 93}, no. 12, 124030 (2016)
  doi:10.1103/PhysRevD.93.124030
  [arXiv:1512.05305 [hep-th]].\\
  B.~Holdom and J.~Ren,
  ``Quadratic gravity: from weak to strong,''
  arXiv:1605.05006 [hep-th].

\bibitem{Salvio:2014soa}
  A.~Salvio and A.~Strumia,
  ``Agravity,''
  JHEP {\bf 1406}, 080 (2014)
  doi:10.1007/JHEP06(2014)080
  [arXiv:1403.4226 [hep-ph]].

\bibitem{Einhorn:2014gfa}
  M.~B.~Einhorn and D.~R.~T.~Jones,
  ``Naturalness and Dimensional Transmutation in Classically Scale-Invariant Gravity,''
  JHEP {\bf 1503}, 047 (2015)
  doi:10.1007/JHEP03(2015)047
  [arXiv:1410.8513 [hep-th]].\\
 T.~Jones and M.~Einhorn,
  ``Quantum Gravity and Dimensional Transmutation,''
  PoS PLANCK {\bf 2015}, 061 (2015).




\bibitem{Mannheim:2011ds}
  P.~D.~Mannheim,
  ``Making the Case for Conformal Gravity,''
  Found.\ Phys.\  {\bf 42}, 388 (2012)
  doi:10.1007/s10701-011-9608-6
  [arXiv:1101.2186 [hep-th]].

\bibitem{Ivanov:1981wn}
  E.~A.~Ivanov and J.~Niederle,
  ``Gauge Formulation of Gravitation Theories. 1. The Poincare, De Sitter and Conformal Cases,''
  Phys.\ Rev.\ D {\bf 25}, 976 (1982).
  doi:10.1103/PhysRevD.25.976
\bibitem{Wheeler:2013ora}
  J.~T.~Wheeler,
  ``Weyl gravity as general relativity,''
  Phys.\ Rev.\ D {\bf 90}, no. 2, 025027 (2014)
  doi:10.1103/PhysRevD.90.025027
  [arXiv:1310.0526 [gr-qc]].

\bibitem{Iorio:1996ad}
  A.~Iorio, L.~O'Raifeartaigh, I.~Sachs and C.~Wiesendanger,
  ``Weyl gauging and conformal invariance,''
  Nucl.\ Phys.\ B {\bf 495}, 433 (1997)
  doi:10.1016/S0550-3213(97)00190-9
  [hep-th/9607110].

\bibitem{Englert:1976ep}
  F.~Englert, C.~Truffin and R.~Gastmans,
  ``Conformal Invariance in Quantum Gravity,''
  Nucl.\ Phys.\ B {\bf 117}, 407 (1976).
  doi:10.1016/0550-3213(76)90406-5

\bibitem{'THooft:2015skl}
  G.~'T Hooft,
  ``Local conformal symmetry: The missing symmetry component for space and time,''
  Int.\ J.\ Mod.\ Phys.\ D {\bf 24}, no. 12, 1543001 (2015).
  doi:10.1142/S0218271815430014


\bibitem{Hooft:2015rdz}
  G.~'t Hooft,
  ``Singularities, horizons, firewalls, and local conformal symmetry,''
  arXiv:1511.04427 [gr-qc].
\bibitem{Blas:2011ac}
  D.~Blas, M.~Shaposhnikov and D.~Zenhausern,
  ``Scale-invariant alternatives to general relativity,''
  Phys.\ Rev.\ D {\bf 84}, 044001 (2011)
  doi:10.1103/PhysRevD.84.044001
  [arXiv:1104.1392 [hep-th]].

\bibitem{Karananas:2016ltn}
  G.~K.~Karananas,
  ``Poincar\'e, Scale and Conformal Symmetries: Gauge Perspective and Cosmological Ramifications,''
  arXiv:1608.08451 [hep-th].


\bibitem{Freedman:1976xh}
  D.~Z.~Freedman, P.~van Nieuwenhuizen and S.~Ferrara,
  ``Progress Toward a Theory of Supergravity,''
  Phys.\ Rev.\ D {\bf 13}, 3214 (1976).
  doi:10.1103/PhysRevD.13.3214












\bibitem{Deser:1976yx}
  S.~Deser, M.~J.~Duff and C.~J.~Isham,
  ``Nonlocal Conformal Anomalies,''
  Nucl.\ Phys.\ B {\bf 111}, 45 (1976).
  doi:10.1016/0550-3213(76)90480-6
\bibitem{Erdmenger:1996yc}
  J.~Erdmenger and H.~Osborn,
  ``Conserved currents and the energy momentum tensor in conformally invariant theories for general dimensions,''
  Nucl.\ Phys.\ B {\bf 483}, 431 (1997)
  doi:10.1016/S0550-3213(96)00545-7
  [hep-th/9605009].



\bibitem{Riegert}
  R.~J.~Riegert,
  ``A Nonlocal Action for the Trace Anomaly,''
  Phys.\ Lett.\ B {\bf 134}, 56 (1984).
  doi:10.1016/0370-2693(84)90983-3

\bibitem{Donoghue:2015xla}
  J.~F.~Donoghue and B.~K.~El-Menoufi,
  ``QED trace anomaly, non-local Lagrangians and quantum Equivalence Principle violations,''
  JHEP {\bf 1505}, 118 (2015)
  doi:10.1007/JHEP05(2015)118
  [arXiv:1503.06099 [hep-th]].\\
J.~F.~Donoghue and B.~K.~El-Menoufi,
  ``Covariant non-local action for massless QED and the curvature expansion,''
  JHEP {\bf 1510}, 044 (2015)
  doi:10.1007/JHEP10(2015)044
  [arXiv:1507.06321 [hep-th]].

\bibitem{Barvinsky:1994cg}
  A.~O.~Barvinsky, Y.~V.~Gusev, G.~A.~Vilkovisky and V.~V.~Zhytnikov,
  ``The One loop effective action and trace anomaly in four-dimensions,''
  Nucl.\ Phys.\ B {\bf 439}, 561 (1995)
  doi:10.1016/0550-3213(94)00585-3
  [hep-th/9404187].

\bibitem{Unruh:1988in}
  W.~G.~Unruh,
  ``A Unimodular Theory of Canonical Quantum Gravity,''
  Phys.\ Rev.\ D {\bf 40}, 1048 (1989).
  doi:10.1103/PhysRevD.40.1048

\bibitem{Ng:1990xz}
  Y.~J.~Ng and H.~van Dam,
  ``Unimodular Theory of Gravity and the Cosmological Constant,''
  J.\ Math.\ Phys.\  {\bf 32}, 1337 (1991).
  doi:10.1063/1.529283

\bibitem{Weinberg:1988cp}
  S.~Weinberg,
  ``The Cosmological Constant Problem,''
  Rev.\ Mod.\ Phys.\  {\bf 61}, 1 (1989).
  doi:10.1103/RevModPhys.61.1


\bibitem{Krasnov:2011pp}
  K.~Krasnov,
  ``Pure Connection Action Principle for General Relativity,''
  Phys.\ Rev.\ Lett.\  {\bf 106}, 251103 (2011)
  doi:10.1103/PhysRevLett.106.251103
  [arXiv:1103.4498 [gr-qc]].




\bibitem{Delfino:2012aj}
  G.~Delfino, K.~Krasnov and C.~Scarinci,
  ``Pure connection formalism for gravity: Feynman rules and the graviton-graviton scattering,''
  JHEP {\bf 1503}, 119 (2015)
  doi:10.1007/JHEP03(2015)119
  [arXiv:1210.6215 [hep-th]].

\bibitem{Basile:2015jjd}
  T.~Basile, X.~Bekaert and N.~Boulanger,
  ``Note about a pure spin-connection formulation of general relativity and spin-2 duality in (A)dS,''
  Phys.\ Rev.\ D {\bf 93}, no. 12, 124047 (2016)
  doi:10.1103/PhysRevD.93.124047
  [arXiv:1512.09060 [hep-th]].


\bibitem{Herfray:2015fpa}
  Y.~Herfray, K.~Krasnov and Y.~Shtanov,
  ``Anisotropic singularities in chiral modified gravity,''
  doi:10.1088/0264-9381/33/23/235001
  arXiv:1510.05820 [gr-qc].

\end{thebibliography}
\end{document}